\documentclass[conference]{IEEEtran}
\IEEEoverridecommandlockouts
\usepackage{cite}
\usepackage{amsmath,amssymb,amsfonts}
\usepackage{algorithmic}
\usepackage{graphicx}
\usepackage{textcomp}
\usepackage{xcolor}
\def\BibTeX{{\rm B\kern-.05em{\sc i\kern-.025em b}\kern-.08em
    T\kern-.1667em\lower.7ex\hbox{E}\kern-.125emX}}
\begin{document}

\title{Role of the Pretraining and the Adaptation data sizes for low-resource real-time MRI video segmentation}


\author{\IEEEauthorblockN{Masoud Thajudeen Tholan, Vinayaka Hegde, Chetan Sharma, Prasanta Kumar Ghosh}
\IEEEauthorblockA{\textit{Department of Electrical Engineering} \\
\textit{Indian Institute of Science}\\
Bengaluru, India \\
masoudt@iisc.ac.in, vinayakahegde619@gmail.com, chetansharma@iisc.ac.in, prasantg@iisc.ac.in}


}

\maketitle

\begin{abstract}

Real-time Magnetic Resonance Imaging (rtMRI) is frequently used in speech production studies as it provides a complete view of the vocal tract during articulation.
This study investigates the effectiveness of rtMRI in analyzing vocal tract movements by employing the SegNet and UNet models for
Air-Tissue Boundary (ATB)
segmentation tasks. 
We conducted pretraining of a few base models using increasing numbers of subjects and videos, to assess performance on two datasets.
First, consisting of unseen subjects with unseen videos from the same data source, 
achieving 0.33\% and 0.91\% (Pixel-wise Classification Accuracy (PCA) and Dice Coefficient respectively) better than its matched condition.
Second, comprising unseen videos from a new data source, where we obtained an accuracy of 99.63\% and 98.09\% (PCA and Dice Coefficient respectively) of its matched condition performance.
Here, matched condition performance refers to the performance of a model trained only on the test subjects which was set as a benchmark for the other models.
Our findings highlight the significance of fine-tuning and adapting models with limited data. 
Notably, we demonstrated that effective model adaptation can be achieved with as few as 15 rtMRI frames from any new dataset.

\end{abstract}

\begin{IEEEkeywords}
Real-Time Magnetic Resonance Imaging, Air-Tissue Boundary Segmentation, SegNet, U-Net, Fine-tuning

\end{IEEEkeywords}

\section{Introduction}


Real-time Magnetic Resonance Imaging (rtMRI) is widely used in speech science and linguistics for studying the dynamics of speech production across different languages and health conditions\cite{hagedorn2019engineering}. Precise segmentation of Air-Tissue Boundaries
(ATBs) in rtMRI is crucial for advancing research, specially in understanding vocal tract dynamics during speech production\cite{bresch2008seeing}. 
The research focused on analyzing vocal tract movements \cite{ramanarayanan2013investigation} and its morphological structure\cite{lammert2013interspeaker} typically employs ATB segmentation as an essential pre-processing step when working with rtMRI video frames.
Various applications of ATB segmentation include tasks such as speaker identification \cite{prasad2015estimation}, text-to-speech synthesis \cite{toutios2016articulatory},
visual augmentation for articulatory videos \cite{chandana2018automatic},
etc. Accurate ATB segmentation of the upper airway is vital for speech processing applications [\citen{parrell2014interaction}, \citen{hsieh2013pharyngeal}] analyzing the dynamic changes in the vocal tract's cross-sectional area over time\cite{story1996vocal} and estimating speech rate from articulatory features \cite{mannem2019acoustic}.
Therefore, obtaining accurate ATB segmentation in each frame of rtMRI videos is essential for understanding the articulatory dynamics in speech production
[\citen{li2016speaker}, \citen{hsieh2013truncation}, \citen{udupa2023real}].


ATB segmentation in rtMRI frames has been extensively explored in past research, employing both supervised and unsupervised methodologies. Supervised approaches achieve higher ATB accuracy but face challenges in generalizing across unseen subjects due to morphological variability. Robust ATB prediction methods using composite analysis grids [\citen{kim2014enhanced}, \citen{ohman1967numerical}, \citen{maeda1979modele}, \citen{proctor2010rapid}] and unsupervised, semiautomatic approaches [\citen{asadiabadi2017vocal}, \citen{lammert2010data}] each offer distinct advantages. Somandepalli et al. \cite{somandepalli2017semantic} introduced deep learning based semantic ATB prediction, which was further advanced by Valliappan et al.\cite{valliappan2018air} through the use of a Fully Convolutional Network (FCN). Advait et al. \cite{koparkar2018supervised} developed Fisher-discriminant measure (FDM) based method for predicting ATBs in test rtMRI images, where the prediction is made by combining ATBs from the training set to maximize the FDM objective function. However, this approach results in a non-smooth predicted ATB, and its dynamics are constrained by the training set. To overcome these limitations, Valliappan et al. \cite{valliappan2019improved} proposed a semantic segmentation-based approach using a 2-D deep convolutional encoder-decoder network (SegNet) for ATB prediction. 
The SegNet based approach, as shown in \cite{valliappan2019improved} and \cite{mannem2019segnet}, provides the best performance in seen subject conditions, and has also demonstrated improved results compared to other 2D-CNN based methods, such as those in [\citen{valliappan2018air}, \citen{mannem2019air}].
Several studies [\citen{valliappan2018air}, \citen{koparkar2018supervised}, \citen{valliappan2019improved}] utilized 8 videos from each subject (F1, F2, M1, M2) from the USC-TIMIT database \cite{narayanan2014real} as training data for ATB segmentation. Additionally, in \cite{valliappan2019improved}, Valliappan et al. also explored the impact of varying the training data size, using 1 to 8 videos per subject to determine the minimum number of rtMRI videos required to achieve saturation.
Renuka et al. \cite{mannem2019air} 
for unseen-subject condition, the model was trained using 16 videos each from three subjects, with 16 videos from one subject reserved for testing. Notably, the approach achieved strong performance with only 30 unseen subject images as adaptation data. However, it focused only on upper and lower contours, omitting the side contour and used data only from four subjects (limiting diversity), and generalized its results to a fixed database with consistent resolution and dimensions, restricting its broader applicability.
These works reveal the role of pretraining datasize, often using multiple videos per subject from a specific database like USC-TIMIT, but they lack diversity and generalizability to new databases having different specifications, such as resolution or frame rate. Furthermore, the labor-intensive annotation of rtMRI data limits scalability. Addressing this, low-resource adaptation becomes essential, enabling robust performance with minimal data. By reducing dependency on extensive annotations, we like to generalize across diverse corpora, making ATB segmentation techniques more efficient and applicable in dynamic and resource-constrained scenarios.

Ribeiro et al. \cite{ribeiro2024automatic} applied Mask R-CNN and a transfer learning approach to estimate the shapes of nine non-rigid vocal tract articulators using a small dataset. Their method demonstrated good generalization to new subjects, requiring only 10 images for adaptation if the subject's position differs from the training set. However, their model failed to predict the contact between articulators, and it only covered the soft articulators. Furthermore, their work relies on static vowel articulation data, limiting its applicability to continuous speech production. It remains unclear how well it adapts to a diverse speaker set given the anatomical variation in articulators. 
In contrast, our models segment both rigid and non-rigid vocal tract articulators, extending to rigid bodies for complete vocal tract shapes, and utilize dynamic MRI data capturing continuous speech. 

In this paper, we employ SegNet\cite{valliappan2019improved} and U-Net\cite{ronneberger2015u} for ATB prediction in the midsagittal plane of rtMRI videos, segmenting both rigid and non-rigid vocal tract articulators using various training-to-validation splits and fine-tuning strategies. We investigate whether fine-tuning models with a limited amount of data from unseen subjects can bridge the performance gap between models trained exclusively on the test subjects and those trained on different subjects. To quantify this, we establish a best-case performance benchmark for each test dataset by training models solely on videos of the test subjects and comparing their performance with fine-tuned models.

\section{Dataset}
\begin{figure}[htbp]
    \centering
    \includegraphics[width=0.5\textwidth]{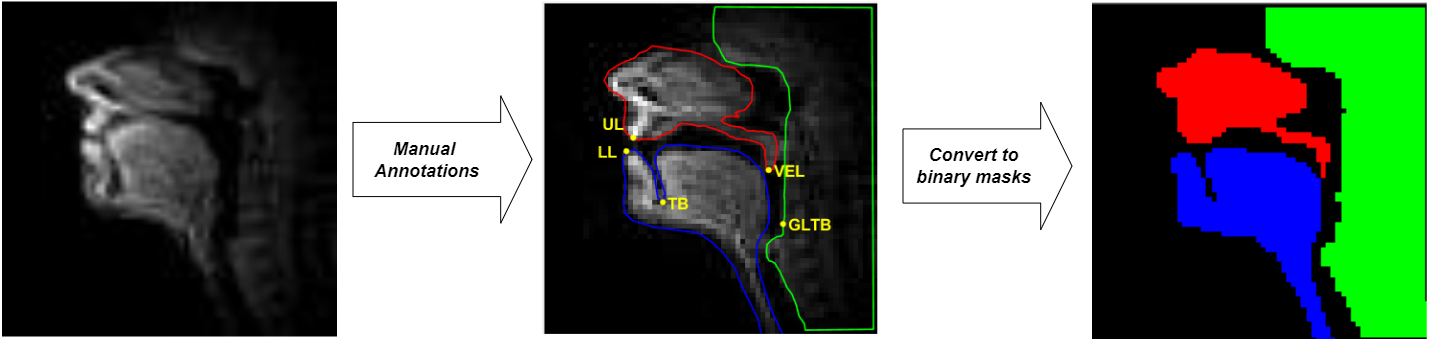}
    \caption{Illustration of rtMRI frame, manual annotations performed, and the conversion to binary masks}
    \label{fig:1}
\end{figure}

We utilize two different datasets in this work. The first one is the USC-TIMIT database\cite{narayanan2014real} consisting of rtMRI videos of the upper airway in the midsagittal plane, for five female (F1, F2, F3, F4, and F5) and five male (M1, M2, M3, M4, and M5) subjects, each speaking 460 sentences from the MOCHA-TIMIT database\cite{wrench2000multichannel}. The videos have a frame rate of 23.18 fps and each frame of the video has a spatial resolution of 68 $\times$ 68 pixels (2.9 $\mathrm{mm}$ $\times$ 2.9 $\mathrm{mm}$). A total of 150 videos were selected, 15 videos from each of the 10 subjects (one for each sentence). The selected videos have 5791 frames for the five female subjects and 6136 frames for the five male subjects. A MATLAB-based Graphical User Interface (GUI)\cite{pattem2018optimal} was used to manually annotate ATBs in each rtMRI frame. Annotations were made for three contours (C1, C2, and C3), and five important anatomical locations were also marked: Glottis-Beginning (GLTB), Tongue Base (TB), Velum (VEL), Lower Lip (LL), and Upper Lip (UL) as shown in Figure \ref{fig:1}. 

The second dataset is a subset of the USC 75-speaker speech rtMRI video database\cite{lim2021multispeaker} in which videos were recorded at a frame rate of 83.277 fps and each frame of the video has a spatial resolution of 84 $\times$ 84 pixels (2.4 $\mathrm{mm}$ $\times$ 2.4 $\mathrm{mm}$).
The rtMRI videos in the given database consist of various Vowel-Consonant-Vowel (VCV) sequences and sentences, where one video may contain multiple VCV sequences or sentences.
We selected one video from each of two subjects (one female and one male) labeled F6 and M6, containing multiple sentences.
Using Audacity\cite{Audacity2023}, we marked the start and end points of each sentence by observing both the waveform and spectrogram. Next, we identified the time frame for all sentences and opted to use one out of every four frames, as the frame rate is high and consecutive frames would not significantly differ. This ensures that we capture meaningful variations in articulation during continuous speech.
In this way, the selected videos were divided into shorter video clips, each containing a single sentence. For our analysis, we used two such clips, referred to as video 1 and video 2, from each subject. 
Video 1 contained 91 frames and video 2 contained 69 frames, for both subjects.
The same MATLAB GUI\cite{pattem2018optimal} was used to annotate this dataset, following the exact method described earlier.

Ground truth binary masks were generated from the three contours, where class 0 represents air and class 1 represents tissue. These ground truth masks were utilized to train the SegNet and U-Net models. This is illustrated in Figure \ref{fig:1}.

\section{Methodology}
Both SegNet \cite{valliappan2019improved} and U-Net \cite{ronneberger2015u} architectures used for semantic segmentation follow an encoder-decoder structure to extract and reconstruct spatial information from rtMRI data. In both models, a downsampling process is followed in the encoder to capture spatial features, which are then reconstructed in the decoder stage. Deeper layers capture increasingly abstract features. U-Net preserves high-resolution information by using transpose convolutions and concatenating corresponding feature maps from the encoder during the decoder stage, while SegNet restores spatial details using transpose convolutions and upsampling. Both models generate three binary masks, through three separate decoder channels, each classified into tissue (class 1) or air cavity (class 0) using the softmax activation function, with Binary Cross-Entropy (BCE) loss to optimize the output as depicted in \cite{valliappan2019improved}.


\section{Experimental Setup}

A total of 12 base models were trained each with SegNet and U-Net models using the videos from the first dataset (68 $\times$ 68 pixels). Four different subject groups were used for the training process.

\begin{enumerate}
    \item F1, M1
    \item F1, F2, M1, M2
    \item F1, F2, F3, M1, M2, M3
    \item F1, F2, F3, F4, M1, M2, M3, M4

\end{enumerate}
The first 10 videos of each subject were used for training and validation purposes, and each group was trained using 3 different training-to-validation splits.

\begin{itemize}
    \item \textbf{2:1 split} – 2 videos for training and 1 for validation
    \item \textbf{4:1 split} – 4 videos for training and 1 for validation
    \item \textbf{8:2 split} – 8 videos for training and 2 for validation
\end{itemize}

The models were trained for a maximum of 30 epochs, and early stopping was used to prevent over-fitting.

\subsection{Fine-tuning to adapt for unseen subjects}

To improve performance on unseen subjects of the same dataset, the fine-tuning of base models was performed using varying numbers of frames (1, 5, 10 and 15), repeated for 10 rounds each to account for variability from random frame selection. Frames were selected from video 11 of subjects F5 and M5. And the entire 12th video was used for validation during fine-tuning. Each fine-tuning process was performed for a maximum of 30 epochs, and early stopping was used to prevent overfitting.

\subsection{Fine-tuning to adapt for unseen corpus}
Here, to improve the performance on the new dataset, the 12 pre-trained base models are fine-tuned in the same way using frames from the first video of F6 and M6 (the first 45 frames are taken for fine-tuning out of which the required number of frames will be chosen randomly and remaining 46 frames are taken for validation).

\subsection{Testing the base and the fine-tuned Models}
Now, we have one set of pre-trained base models, which is tested with the test videos from both the datasets and two sets of fine-tuned models, which are tested on the test videos from the datasets whose videos were used for fine-tuning. The test videos for the first dataset consist of videos 13, 14, and 15 from subjects F5 and M5 and for the next dataset, the test set is the second video of F6 and M6. The evaluation uses the Dice Coefficient and the Pixel-wise Classification Accuracy (PCA) metrics, as described in \cite{udupa2023real} and \cite{valliappan2019improved}, respectively. Also, to compare the results obtained, a benchmark was set for each dataset by training a model using the videos from the test subjects of that dataset. The performance of these models on the test sets is referred to as the matched condition performance. For the first dataset, the first eight videos of subjects F5 and M5 were used for training (the next two videos for validation). For the second dataset, the first 70 percent of the first video was used for training (the remaining 30 percent for validation).

\section{Results and Discussion}

\begin{figure}[h]
    \centering
\includegraphics[width=\linewidth]{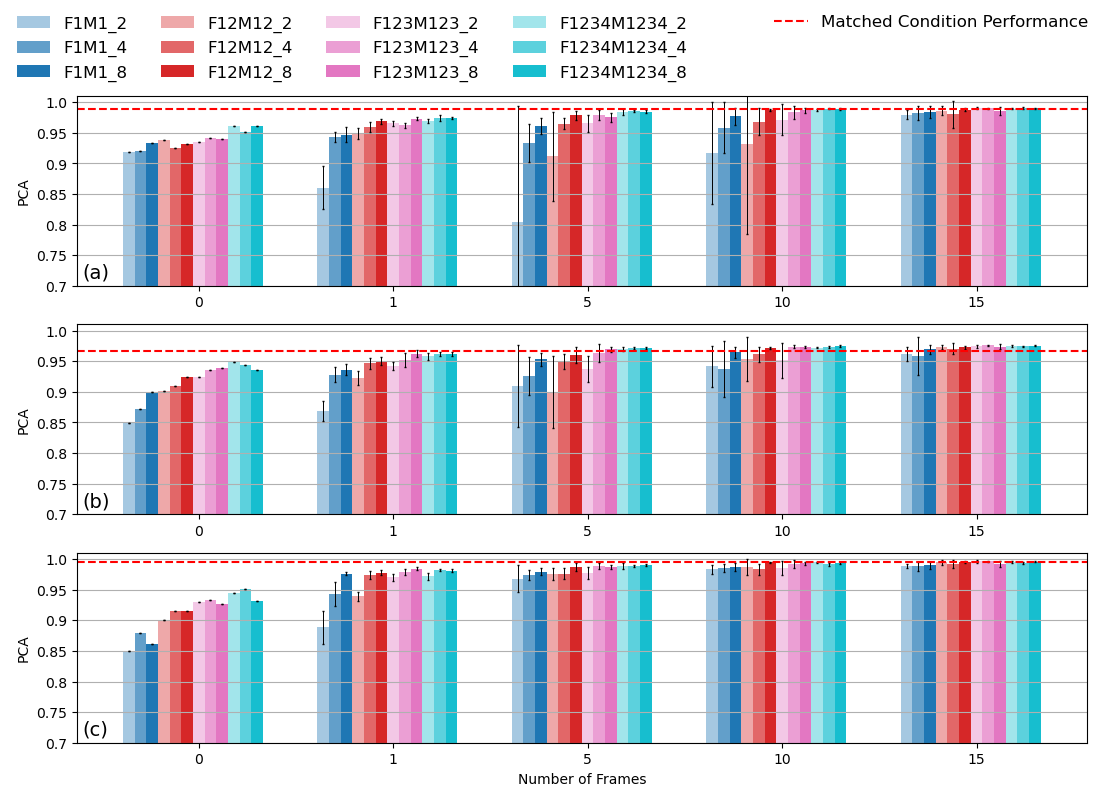}
    \caption{PCA Results for Base Models Fine-Tuned with Varying Frame Counts: Evaluated on F5 and M5 -- (a) Mask 1, (b) Mask 2, (c) Mask 3}
    \label{fig:3}
\end{figure}

\begin{figure}[h]
    \centering
\includegraphics[width=\linewidth]{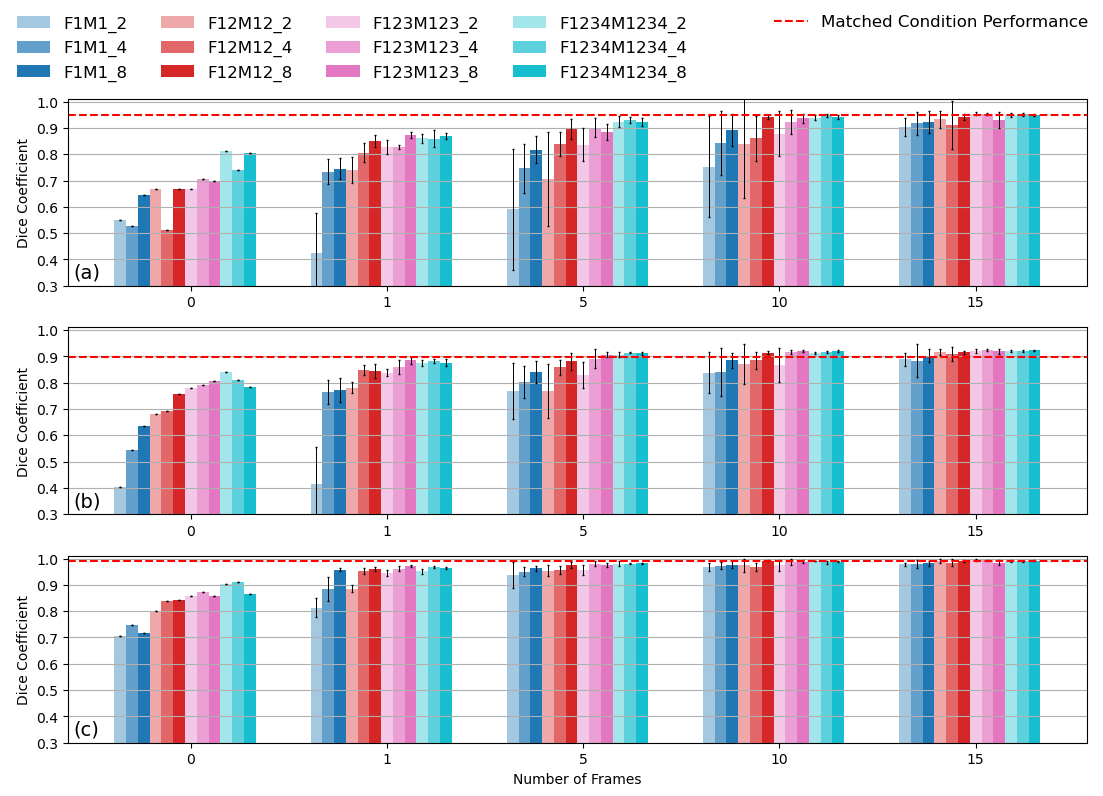}
    \caption{Dice Coefficient Results for Base Models Fine-Tuned with Varying Frame Counts: Evaluated on F5 and M5 -- (a) Mask 1, (b) Mask 2, (c) Mask 3}
    \label{fig:4}
\end{figure}

Figure \ref{fig:3} and Figure \ref{fig:4} represent the PCA and the Dice Coefficient values, respectively, for the unseen subjects F5 and M5. The legends represent the pre-trained models with varying training configurations. For example, ``F1M1\_2'' means the model was trained using subjects F1 and M1 with 2 videos each, and ``F12M12\_4'' indicates training with subjects F1, F2, M1, and M2 using 4 videos each. The increasing sequence of subjects and video numbers, such as in ``F1234M1234\_8,'' highlights progressively broader training datasets. The mean performance of 10 rounds for each number of frames selected is represented by the bars in the figures, and the error lines show their standard deviation.

From the graphs of PCA and Dice Coefficient as shown in Figure \ref{fig:3} and Figure \ref{fig:4}, respectively, obtained by the evaluation of the first dataset, we can observe the following.
\begin{enumerate}

    \item Performance consistently improved as the number of videos used to pre-train the model with subjects F1 and M1 increased. Increasing the number of subjects used during fine-tuning also enhanced performance, highlighting the importance of diverse data in model training.
    \item The error bars representing the standard deviation indicate significant variability in performance, especially when using fewer frames (1 and 5 frames). This suggests that random frame selection greatly influences fine-tuning outcomes when data availability is limited. As the number of frames increases to 10 and 15, the error margins narrow, reflecting more stable and reliable model performance.
    \item The prediction of mask 2 always has the lowest performance among the three masks. This is because the TB regions are more error-prone due to higher motion across the video and more variation compared to other regions as explained in [\citen{roy2022error}, \citen{roy2022air}].
    For such challenging mask 2s, the fine-tuned models began to outperform the matched condition performance, suggesting that the fine-tuning is effective in improving performance, even in challenging areas.

\end{enumerate}

\begin{figure}[h]
    \centering
\includegraphics[width=\linewidth]{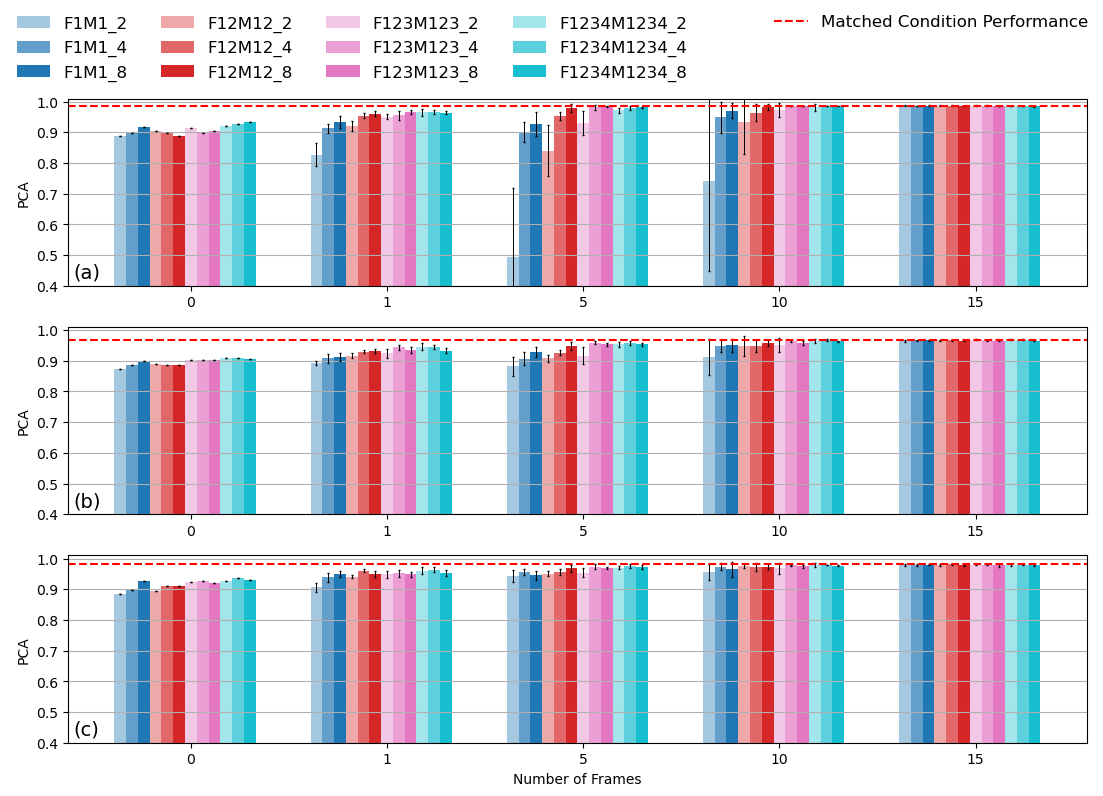}
    \caption{PCA Results for Base Models Fine-Tuned with Varying Frame Counts: Evaluated on F6 and M6 -- (a) Mask 1, (b) Mask 2, (c) Mask 3}
    \label{fig:5}
\end{figure}

\begin{figure}[h]
    \centering
\includegraphics[width=\linewidth]{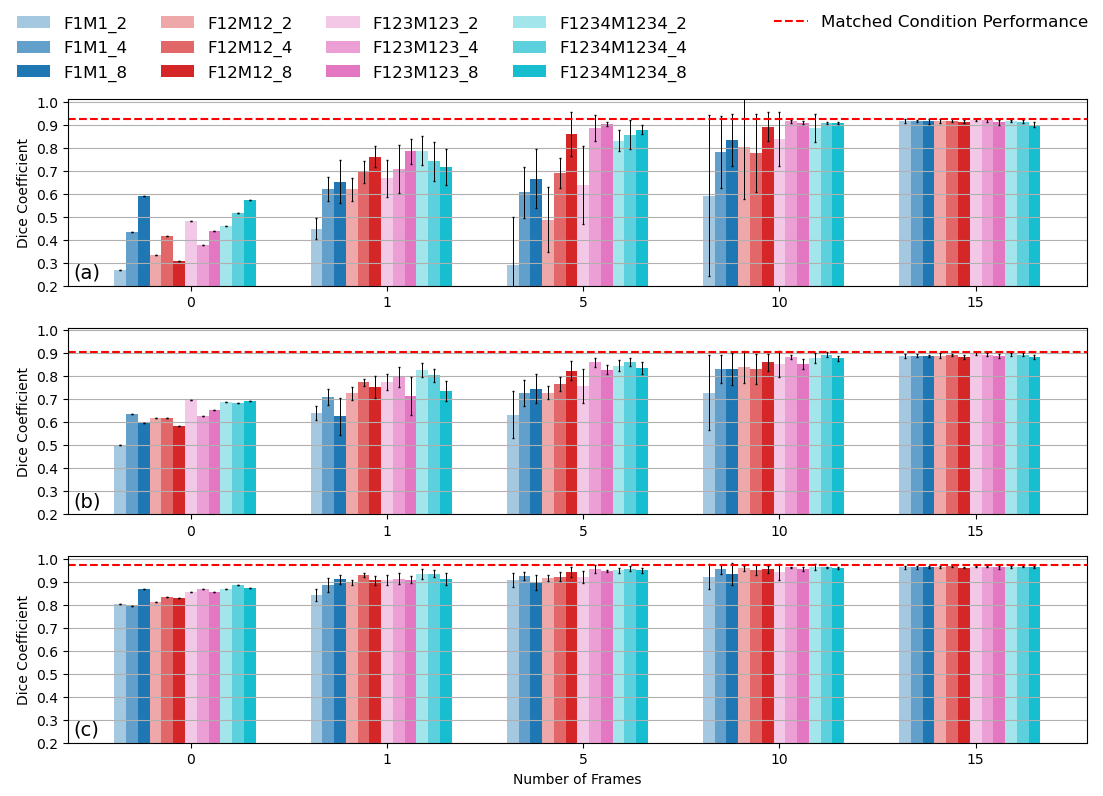}
    \caption{Dice Coefficient Results for Base Models Fine-Tuned with Varying Frame Counts: Evaluated on F6 and M6 -- (a) Mask 1, (b) Mask 2, (c) Mask 3}
    \label{fig:6}
\end{figure}

The graphs obtained from the evaluation of the second dataset are presented in Figure \ref{fig:5} and Figure \ref{fig:6}. These graphs show similar observations, except that the performance in predicting mask 2 is restricted to the matched condition performance and does not exceed it when there is a change in resolution. Overall, these findings emphasize the effectiveness of frame-based fine-tuning in adapting models to unseen subjects. While extensive pre-training with diverse subjects establishes a strong foundation, fine-tuning with a sufficient number of frames can bridge the performance gap. This approach is particularly valuable in scenarios with limited data availability, making it a highly applicable strategy for enhancing model performance in real-world applications.
We found similar observational patterns using the U-Net model as well. 

From the values plotted, the maximum PCA and Dice coefficient values obtained by the base models fine-tuned for the first dataset are 0.33\% and 0.91\% better than the performance of the matched condition, respectively. The maximum performance of the models fine-tuned for the second dataset reached 99.63\% and 98.09\% of the PCA and Dice coefficient values obtained by its matched condition performance, respectively.
Notably, after fine-tuning with 15 frames, the models began to show saturated performance. In addition, even the base models trained on minimal data achieved performance comparable to those trained on larger datasets.
Our approach is also highly effective in addressing technical discrepancies such as resolution differences between pretraining and evaluation datasets.

\section{Conclusion}


In this paper, we investigated the minimum number of frames required for a pretrained model to adapt to unseen subjects for ATB segmentation in rtMRI videos. These unseen subjects may belong to a different corpus with varying specifications, such as spatial resolution and frame rate, compared to the dataset used for pretraining. 
Our results demonstrated that fine-tuning the base models with as few as 15 frames is highly effective in adapting to new subjects, achieving high segmentation accuracy. 
For future work, we plan to explore active learning methods to identify and select frames with greater variations, aiming to further reduce the number of frames needed for fine-tuning while maintaining segmentation performance.

\bibliographystyle{IEEEtran}
\bibliography{mybib}

\end{document}